\documentclass[11pt, a4paper, pdftex]{article}
\usepackage[utf8]{inputenc}
\usepackage[T1]{fontenc}

\usepackage{amsmath}
\usepackage{amssymb}
\usepackage{setspace}
\usepackage{latexsym}
\usepackage{fancyhdr}
\usepackage{lastpage}
\usepackage{times}

\usepackage{graphicx}

\usepackage[english]{babel}

\usepackage{authblk}

\graphicspath{
	{my_figures/}
}

\newcommand{\pa}{\paragraph{}}

\def\og{\leavevmode\raise.3ex\hbox{$\scriptscriptstyle\langle\!\langle$~}}
\def\fg{\leavevmode\raise.3ex\hbox{$\scriptscriptstyle\rangle\!\rangle$}}

\begin{document}

\title{Some clues to build a sound analysis relevant to hearing}
\author[1]{Laurent Millot \thanks{Electronic address: l.millot@ens-louis-lumiere.fr; corresponding author}}
\affil[1]{ENS Louis-Lumière, Noisy-le-Grand, 7 allée du promontoire, 93160, France}

\date{19/07/2005}

\maketitle

	\begin{abstract}
Analysis tools used in research laboratories, for sound synthesis, by musicians or sound engineers can be rather different. Discussion of the assumptions and of the limitations of these tools permits to propose a first tool as relevant and versatile as possible for all the sound actors with a major aim:  one must be able to listen to each element of the analysis because hearing is the final reference tool.  This tool should also be used, in the future, to reinvestigate the definition of sound (or Acoustics) on the basis of some recent works on musical instrument modeling, speech production and loudspeakers design. Audio illustrations will be given.

Paper 6041 presented at the 116th Convention of the Audio Engineering Society, Berlin, 2004
	\end{abstract}

\section{Introduction}
\pa According to the nature of the work on sounds, the analysis tools can be rather different. In sound synthesis lots of specific analysis tools are built according to the assumptions on the nature of the signals: wavelets or frame adaptative basis, noise + sinusoids basis, granular analysis, filter/source analysis, spectrograms, mixed models, ...   According to their style, musicians use waveforms, spectra and spectrograms but mainly their ears. Sound engineers use theirs ears with or without the help of filter banks, the waveforms when a computer based tools is available, sometimes the spectrograms (bargraphs evolving with temporal modification of the energy of the central frequency). To study perception of complex or real sounds, scientists seem to use mainly the spectrograms. But musicians, sound engineers and some of the perception scientists argue that the available tools do no work like the human perception and that hea\-ring stays the reference. And, as acousticians are mostly working with the model of wave propagation, they mainly use spectra to study transfert functions of linear filters (impedance, admittance, reflexion functions, harmonic content), the spectrograms (evolution of the harmonic content) and the waveforms (pressure signals, impulse or green response, ...).  You can even find some scientists studying acoustics or, even worse, musical acoustics which never try to listen to the sound of their equations...

\pa In this paper we present some clues to derive a sound analysis which may be relevant according to the listening. Indeed, we consider that hearing is the final reference tool when working on audio: if, for instance, the physical (or mathematical) model for an acoustic phenomenon is relevant then it may sound similar to the real acoustic phenomenon. To achieve this aim, we introduce the following constraints to build  the analysis tools:
\begin{itemize}
\item listening of  every component of the analysis;
\item no distorsion introduced by the analysis;
\item independent modification of every component with classical tools (digital filters or  audio effects);
\item choice to  remix (or synthesize) every desired selection of the analysis components;
\item components as relevant as possible to the human perception and to the "production habits" of musicians and of sound engineers;
\item  technical principles as simple as possible to permit a real-time implementation in the future. 
\end{itemize}

\pa The organisation of the paper is then the following. In section 2, we review the available tools for sound analysis and finally show why we need to derive one based on the à trous algorithm. In section 3, we explain the main principles involved in the algorithm we used, give the different frequency mappings which are available in our programs and review the ideas to display useful information. Section 3 discusses the opportunity to choose one frequency mapping, the possible extensions or modifications of the current  tools and proposes some first experiments to perform with these tools. 

\section{Choice of the analysis principle}
\subsection{Waveforms}
\pa The first analysis tool is the waveform of the studied sound. In figure 1 two waveforms are presented: spoken french message "Base du traitement du signal" on the top; a pressure measurement inside a diatonic harmonica while performing a G blow on the fourth channel. These signal are used all over the paper to illustrate the different analysis.

\begin{figure}
\begin{center}
\includegraphics[width=7.5cm]{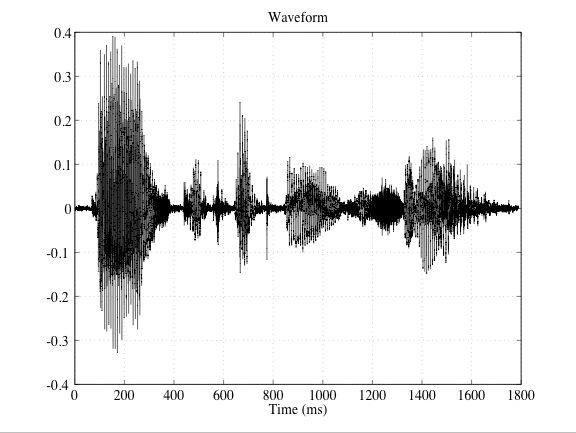} 
\includegraphics[width=7.5cm]{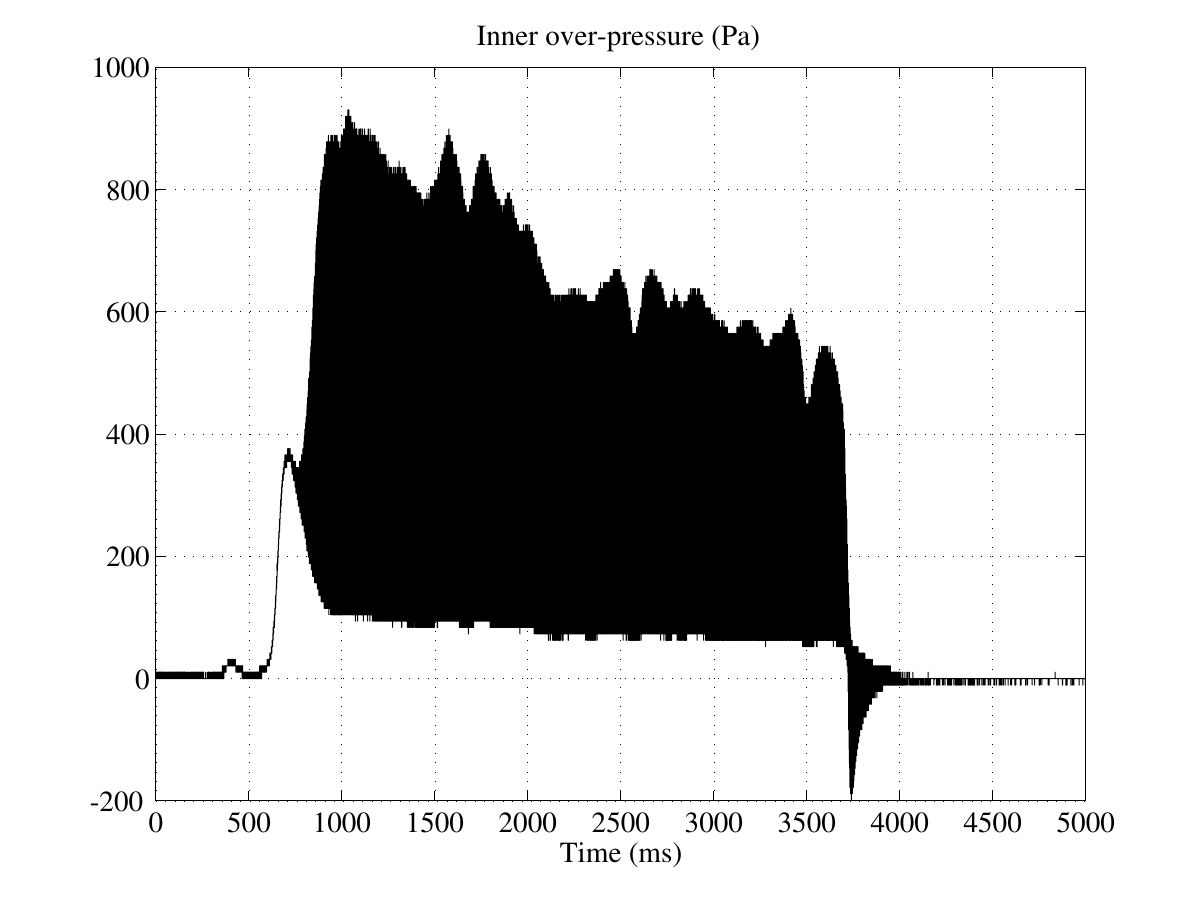} 
\caption{examples of waveforms: (left) spoken message in french, (right) over-pressure measurement for a normal G blow (fourth channel) on a G diatonic harmonica.}
\end{center}
\end{figure} 

\pa By considering the whole waveforms, some preliminary remarks can be made. One can note the temporal segmentation of the spoken message and is then able to extract the different words. It is also possible to note that the level has great variation over a zero mean value: the signal is clearly bipolar as lots of classical sounds. The pressure signal (right) is unipolar which means the (mathematic) mean is non zero, that the magnitude is quite great (a mean level higher than 140 dB SPL) but the periodic behavior of the sound is not discernable without a zoom. With the zoom on a small part of the pressure signal (figure 2), the quasi periodic nature of the sound becomes obvious and as the temporal variations are fast, like some shock waves, one can understand that the mean level has only a mathematical sense and that the harmonic content must be extended. But it is difficult to determine this harmonic content and its precise temporal evolution with only the waveforms. For this task, the experienced ears are a rather better tool but to detect a clic, waveforms can constitue a convenient tool. Even if the waveform gives access to the temporal behavior of the signal values we need another tool to access the frequential content.
 
 \begin{figure}
\begin{center}
\includegraphics[height=7.5cm]{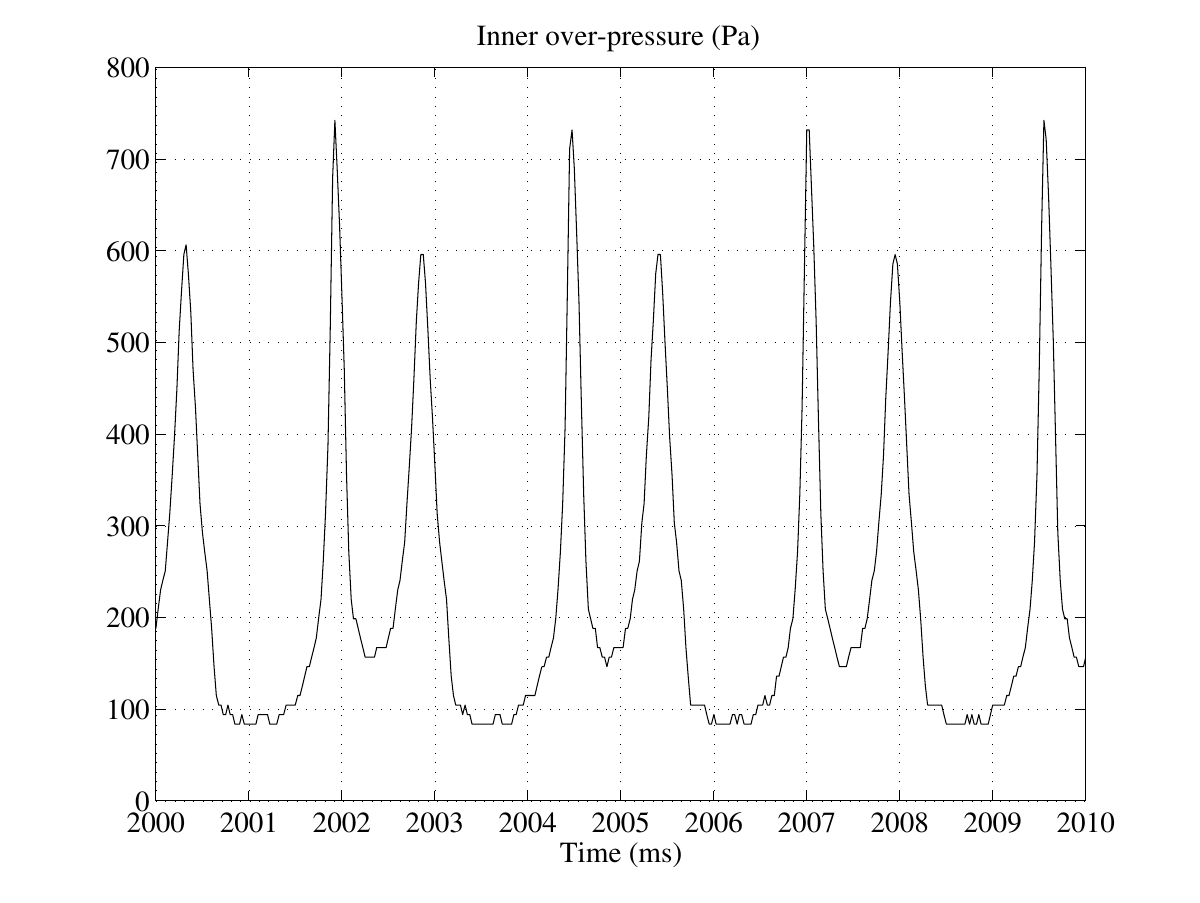} 
\caption{zoom to point out a few "periods" of the over-pressure signal for the former blown note on a diatonic harmonica.}
\end{center}
\end{figure} 

\subsection{Spectra and spectrograms}
\pa Another kind of classical analysis tools are built on the Fourier transform \cite{Mallat} : the fast Fourier transform which gives the spectra of the signal (magnitude of the transform) ; the short time Fourier transform which gives a succession of spectra over a sliding short term window, the spectrogram. Figures 3 and 4 respectively show the spectra of both test signals and the related spectrograms. 

\pa The spectra illustrate the relative weight of each calculated frequency for the whole sound: we do not have any information about the times at which these frequencies are present, we know their mean weight. The spectrograms give us some information about how the frequencies vary with time. But, we have to use a short window which means a poor frequency resolution and a distorsion of the spectra as multiplying parts of the signal by a temporal window is equivalent to convolve the spectra of the signal and of the window. This means that we have to realize a compromise between the temporal and frequential precisions when choosing the length and the nature of the window. Moreover, the spectrogram has another drawbacks: the frequential mapping is regular which does not meet the human perception; we can hardly listen to a part of the spectrogram without some spharpy filters but as it calculates the transform for fixed frequencies it could be not really interesting to listen to parts of it. We do not introduce the waterfall as it seems less easy to use than the spectrogram. For our goal, the spectra and the spectrogram are not adapted so we need another tool permitting  a listening and closer to our perception.

 \begin{figure}
\begin{center}
\includegraphics[width=7.5cm]{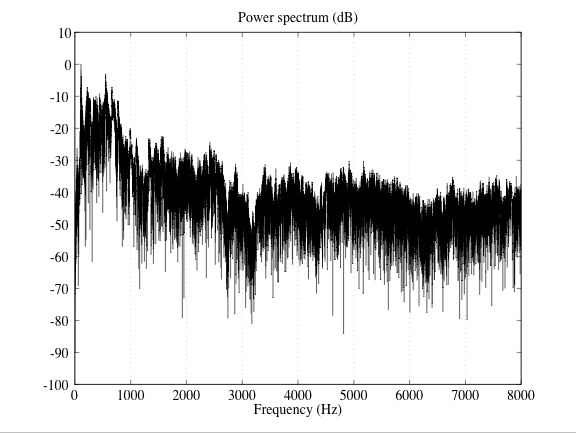} 
\includegraphics[width=7.5cm]{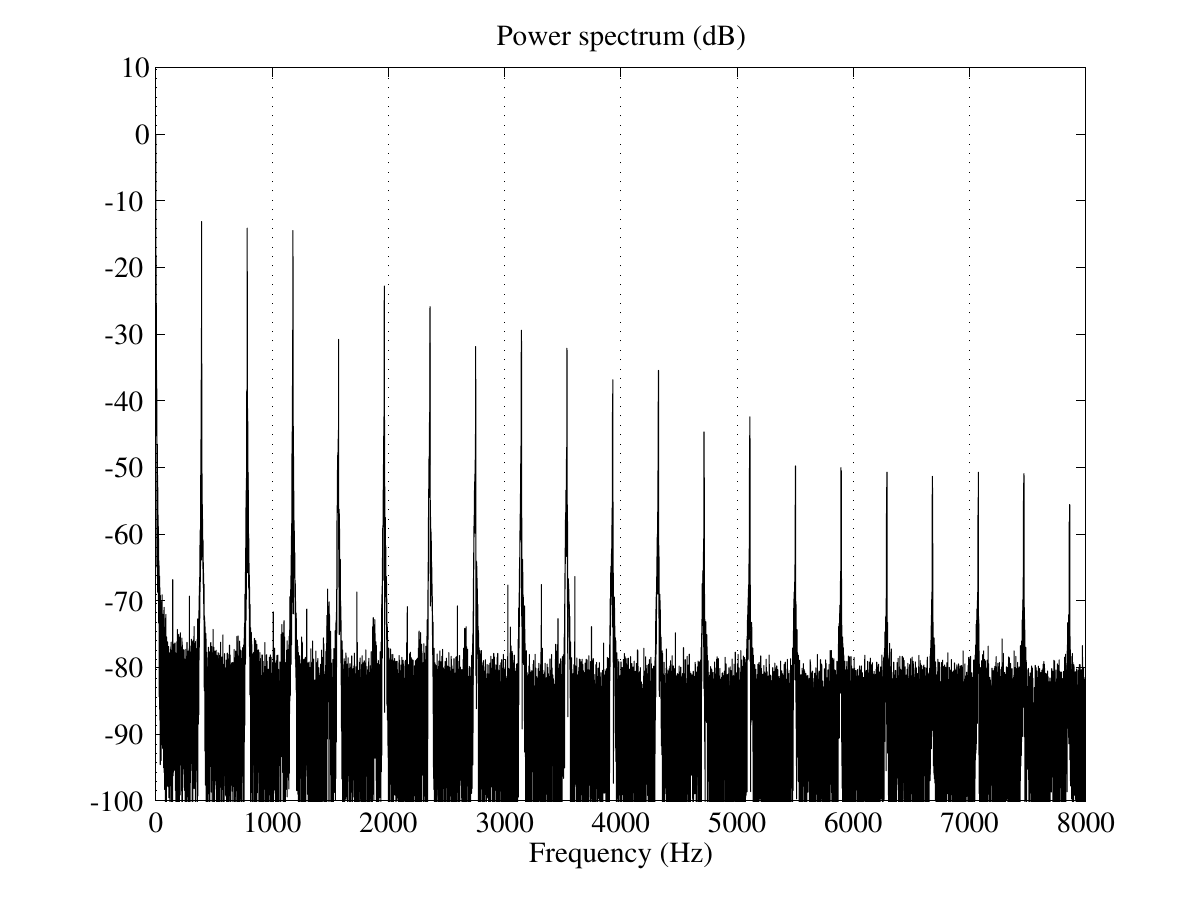} 
\caption{Spectras of both test signals: (left) speech, (right) pressure measurement.}
\end{center}
\end{figure} 

 \begin{figure}
\begin{center}
\includegraphics[width=7.5cm]{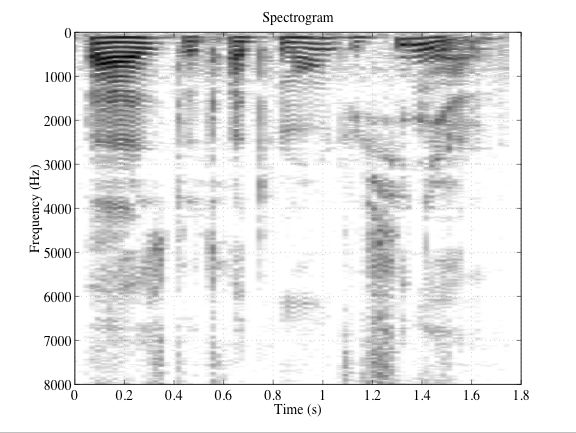} 
\includegraphics[width=7.5cm]{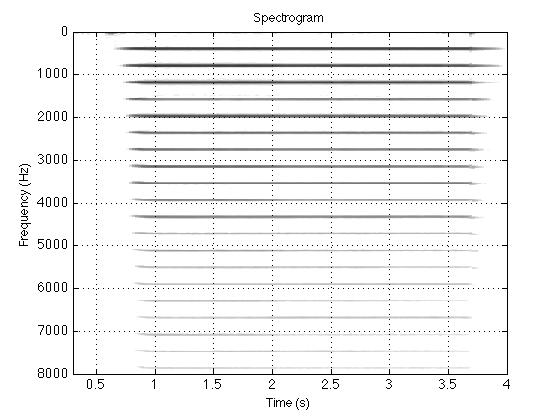} 
\caption{Spectrograms of both test signals using 2048 points FFT: (left) speech, (right) pressure measurement.}
\end{center}
\end{figure}

\subsection{Time-frequency algorithms}
\pa Another tool can be the familly of the wavelets transforms which can use a non regular mapping for the frequencies axis \cite{Mallat}. If we consider only the case of the discrete wavelets transform, we could think to use the fast wavelet transform calculated with the Mallat's algorithm using a dyadic decomposition: the low frequency subband is divided in two equal subbands while it is necessary and/or meaningful.  The analysis and synthesis stages are illustrated on figure 5 and a plot of all the subbands is given in figure 6.

\pa When considering the different subbands plots, we can access to their temporal evolution as the plots are the waveforms of the subbands. But, as within the analysis stage (see figure 5) each decomposition stage is related to a subsampling of factor 2 we loose the time correlation between all the subbbands and introduce aliasing as the decomposition filters are not ideal. By subsampling the subbands for each new analysis stage, we also change the sampling frequency of each subband which means that if we listen to these subbands we have to adapt the sampling frequency of the sound player. A solution, to keep the initial playing frequency, is to resynthesize all the subbands with the synthesis algorithm by putting all the unwanted subbands to zero. This trick gives subbands which can be listened to but it does not suppress the aliasing, which can become quite noticeable... Moreover, if we want to modify the subbands by different filterings, we still get some aliasing as we do not apply exactly the same filtering to all the transition subbands (where aliasing is hidden). The Mallat like algorithms which use subsampling to keep the same low and high frequency filters for the successive analysis stages are not the right solution. We can not afford subsampling, because of aliasing, according to our rules.

\begin{figure}
\begin{center}
\includegraphics[width=12cm]{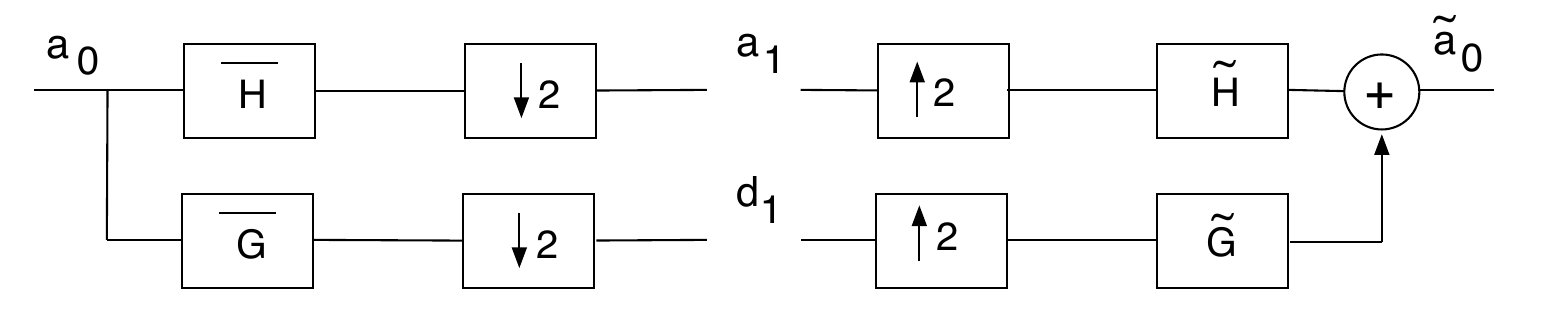} 
\caption{Mallat algorithm for fast dyadic wavelet transform : (left) analysis stage, (right) synthesis stage.}
\end{center}
\end{figure} 

\begin{figure}
\begin{center}
\includegraphics[width=7.5cm]{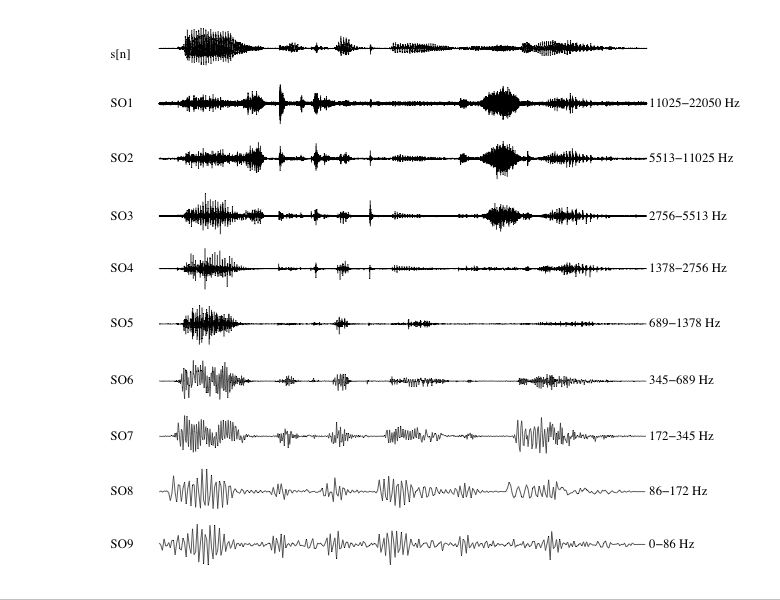} 
\includegraphics[width=7.5cm]{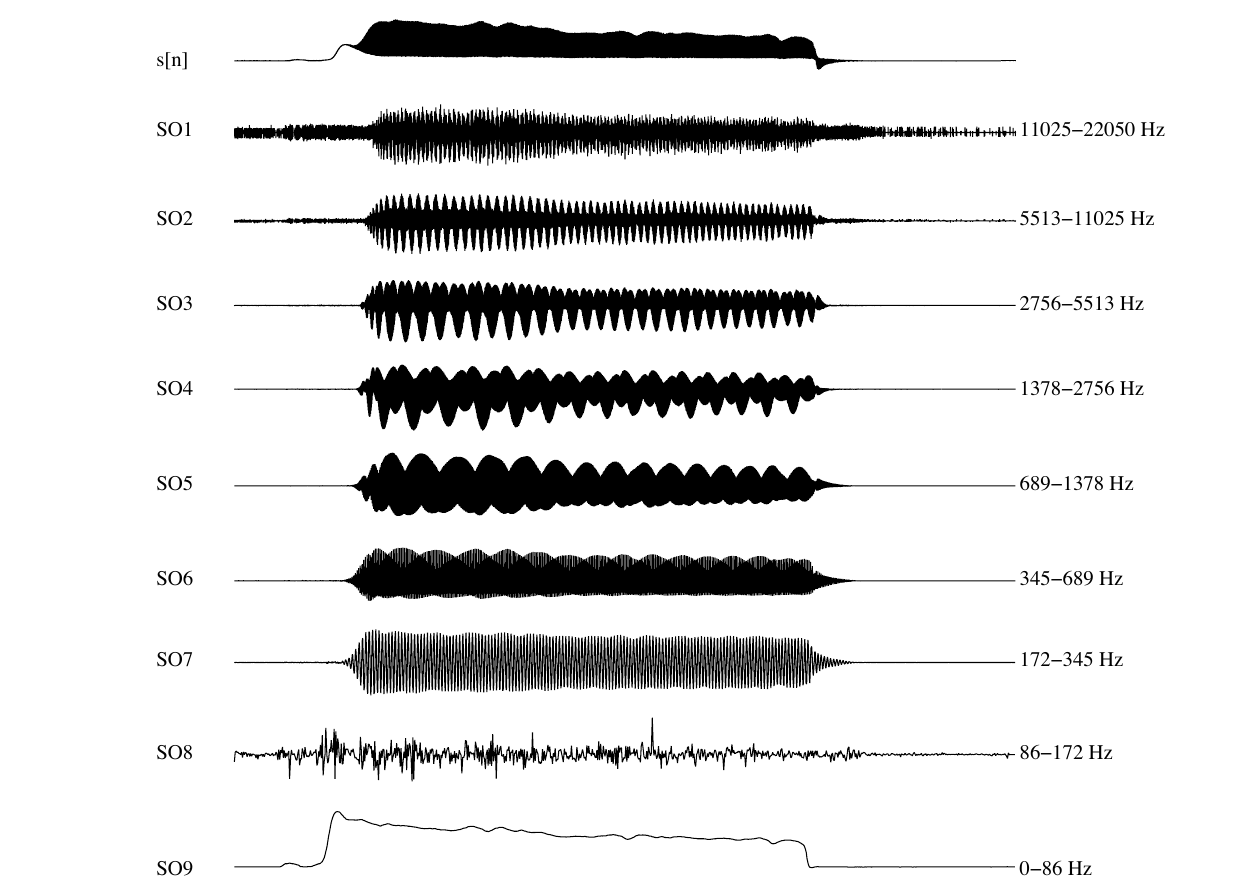} 
\caption{Plots of the subbands for out test signals using the dyadic Mallat's algorithm: (left) speech, (right) pressure measurement.}
\end{center}
\end{figure}

\pa Another algorithm, for the discrete wavelets, is the so called \`{a} trous algorithm \cite{Mallat,  Bijaoui, Holdschneider, Shensa}. With the \`{a} trous algorithm, there is no more subsampling of the signal but an oversampling of the filters for each new stage: the filters for the new analysis stage are derived by inserting one zero between each couple of samples of the impulse response of the former analysis filters; one can think about a temporal dilatation of the decomposition filters for each new analysis stage. By oversampling the analysis filters, each subband has the same sampling frequency which means that there is no aliasing. Moreover, there is no more need to satisfy a cancellation condition for the analysis and synthesis filters and, using only the design condition (the perfect reconstruction condition) we can simplify the design of the analysis/synthesis algorithm:
\begin{itemize}
\item use of only an analysis filter, $h$ for the first stage,  which gives a more crude approximation of the initial signal $a_0[n]$ called $a_1[n]=(h\ast a_0)[n]$ ($\ast$ represent the convolution);
\item calculation of the detail signal $d_1[n]$ by subtracting the new approximation $a_1[n]$ to the former one $a_0[n]$ : $d_1[n]=a_0[n]-a_1[n]$;
\item oversampling of the filter $h$ by  factor 2 and iteration of the whole process to derive new approximations and details while it still makes sense to derive a cruder approximation;
\item the original signal $a_0[n]$, or a modified version of it, are simply derived by a mix (an addition) of all the desired subbands.  
\end{itemize} 

\pa Using the \`{a} trous algorithm with a dyadic frequential  mapping, permits to get subbands signals which can be listened, which do no introduce aliasing and already permit to process independently each subband with classical digital filters and/or audio effects. Each new decomposition stage takes ones signal (old approximation) but gives two signals (new approximation and details) which are longer than the former approximation: filtering adds $N_h-1$ samples to the output if $N_h$ is the length of the filter. So, at the end of the analysis process, we have $N_{st}+1$ audio signals (longer than the original signal), when using $N_{st}$ successive decomposition stages.  This is a major drawback when looking for compression but, for a research project dealing with the analysis of the acoustical phenomena it is not a great problem. We just have to be careful about the programmation of the algorithm (use of the memory for instance). Figure 7 shows the different subbands calculated in the case of the dyadic \`{a} trous algorithm.

  \begin{figure}
\begin{center}
\includegraphics[width=7.5cm]{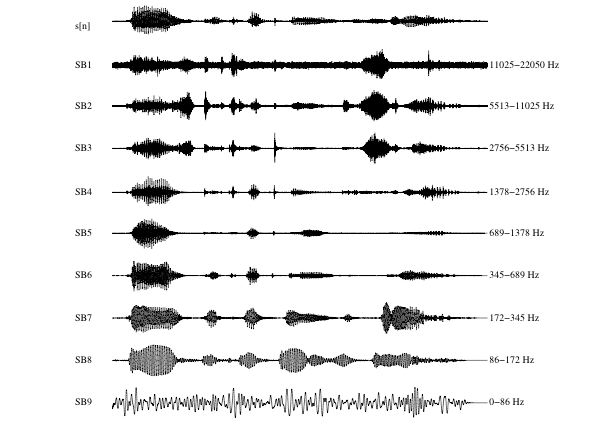} 
\includegraphics[width=7.5cm]{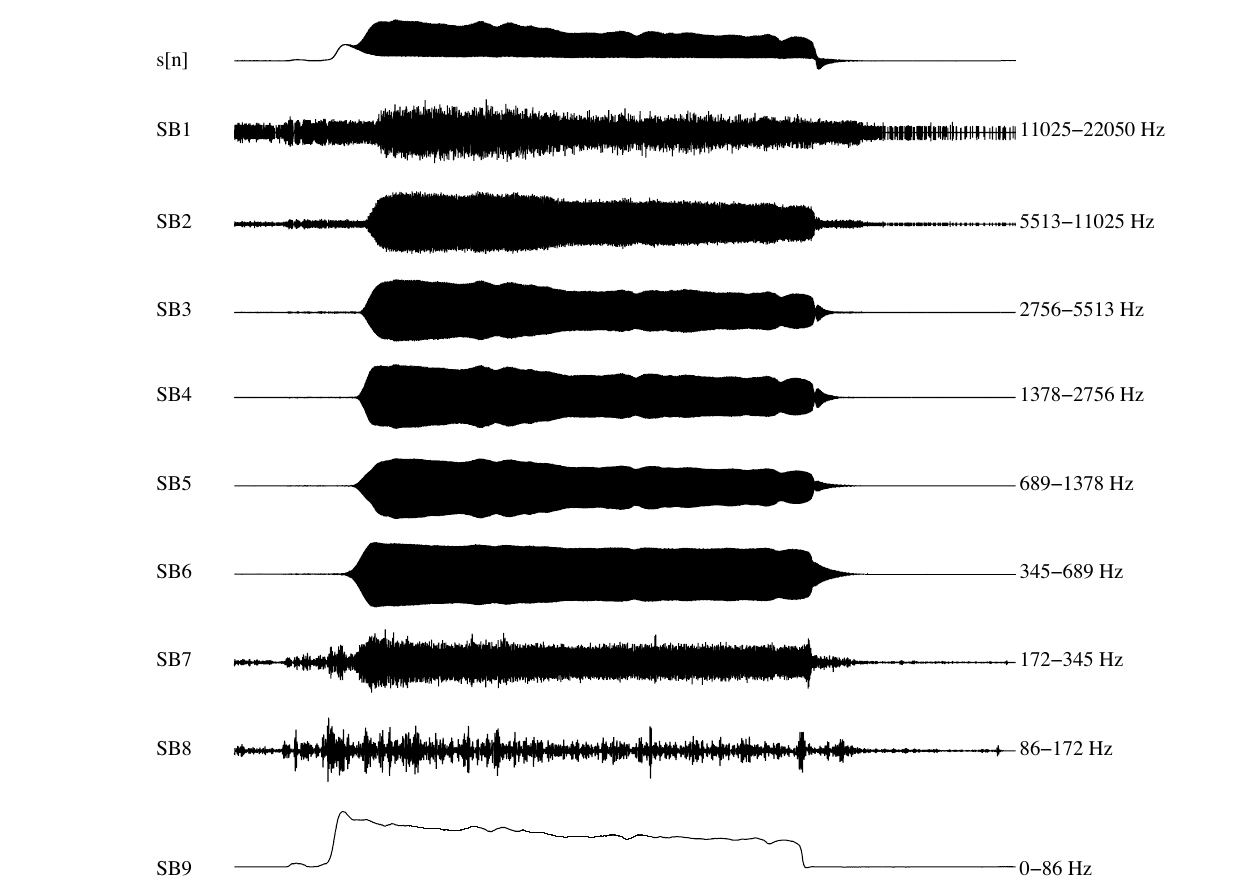} 
\caption{Plots of the subbands for out test signals calculated using the dyadic à trous algorithm: (left) speech, (right) pressure measurement.}
\end{center}
\end{figure} 

\section{Design of the analysis tools}
\subsection{Main algorithms}
\pa For the moment, all the tools are programmed using Scilab 2.7, a free Matlab-like tool for numerical simulation. A port in a fast language (C or C++ for instance) will be done as soon as possible. In the following we give some details about the core of the analysis process in case of monophonic signals but the tools for the stereophonic signals are also available. An extension to multi-channel signals could be done in the future.

\pa The first details to be given concern the decomposition process used for each new stage. In order to not introduce any phase distorsion, we have adopted FIR filters with linear phase. As we do not want to introduce any temporal shift between all the final subbands, we only consider symmetric (or antisymmetric for the high-pass filter case) FIR non-causal filters having an odd number of coefficients. Indeed, to derive the $d_{j}[n]$ detail signal, for stage number $j$, we do not simply calculate $a_{j-1}[n]-a_{j}[n]$  but perform the following operations:
\begin{itemize}
\item calculation of the oversampled version of the new decomposition filter $h_j[n]$ by inserting one zero between each sample of the old decomposition filter $h_{j-1}[n]$ which always gives a filter with an odd number of coefficients;
\item calculation of the group delay for $h_j[n]$ given by $\tau_j= \frac{N_{hj}}{2}$,if $N_{hj}$ is the number of coefficients of $h_{j}[n]$, with $\tau_j$ being an integer number of samples;
\item calculation of the new approximation $a_{j}[n]$ by convolving the old approximation $a_{j-1}[n]$ with $h_j[n]$ (using FFT-fast convolution for instance);
\item  suppression of the first $\tau_j$ coefficients in $a_{j}[n]$ which correspond to the non-causal part of the new approximation;
\item calculation of the detail signal $d_j[n]$ given by the difference between $a_{j-1}[n]$ zero-padded with $\tau$ zeros and the shortcutted $a_j[n]$.
\end{itemize}

\pa For the synthesis, the important point is thinking to zero-pad the shorter subbands in order to add signals having the same length: as we calculate the cruder approximations by filtering we increase the length of the new approximations at each stage; the crudest approximation is then the longest subband. This zero-padding problem may be different for a real-time implementation of the decomposition: it could be necessary to delay the finer details signals if using the successive stages processing. It could be useful to suppress the zeros at the end of the synthesized signal, for instance if no subband additional process is used.    

\pa One may note that the only constraint to satisfy for our analysis filters is to be a FIR and linear phase filters with an odd number of coefficients: one can use any filter which satisfy this condition whether is its width for the transition band. These properties are used in the following to derive some non regular decompositions of the frequencies axis.
 
\subsection{Choice of  frequency mappings}
\pa As introduced in the former section, the first frequential mapping available is the dyadic decomposition, which is useful to point out the introduction of aliasing when using the Mallat's algorithm. But we do not think this mapping is really close to the human perception of sound. So we have introduced, for the moment, two other mappings.

\pa The first  new mapping is an {\it a priori} used  mapping based on an octave decomposition: the 0-16 kHz band is decomposed is as many sub-octaves as needed according to the studied sound and the person which uses the analysis. In this case, the analysis process is based on two original low-pass filters with respectively 8 and 16 kHz cutting frequencies. The analysis algorithm is the following:
\begin{itemize}
\item filtering of the original signal $a_0[n]$ using the 8 kHz low-pass filter which gives a 0-8 kHz approximation signal $a_1[n]$ and an 8 kHz-$\frac{F_s}{2}$ detail signal $d_1[n]$ where $F_s$ is the sampling frequency;
\item using of the iteration process (over-sampling of the low-pass filter, filtering and subtraction) to decompose the 0-8 kHz in as many as needed subbands;
\item decomposition of the 8 kHz-$\frac{F_s}{2}$ detail signal $d_1[n]$ in two subbands, the 8-16 and the 16-$\frac{F_s}{2}$ ones, using the 16 kHz low-pass filter. 
\end{itemize}  

\pa We have not used, firstly, the 16 kHz low-pass filter because of the iterated oversampling process. Indeed, oversampling the decomposition filter is equivalent to a temporal  dilatation of its impulse response which correspond with a compression, by a factor 2 here, of its transfert function. And, adding each new analysis stage is equivalent to convolve the original filter with each new oversampled one. But, the equivalent high-pass and low-pass filters for each stage must realize, according to our choices, a good approximation of the ideal decomposition of a band in two equal sub-octaves. This is only possible if the original filter has a cutting frequency lower than half the Nyquist frequency. As we have used mainly 44100 Hz or 48000 Hz sampled sounds we could not consider the 16 kHz low-pass filter, for the iterative over-sampling process, and have chosen the 8 kHz low-pass filter. With 96 or 192 kHz sampled sounds, it would then be possible to use the 16 kHz low-pass filter as original filter and to not introduce the 8 kHz low-pass filter. It should also be interesting to study if more precision is needed in the 8-16 kHz band and if we should cut this band in two or more subbands (8-12 and 12-16 kHz for instance). But this add-on is not difficult to introduce as we just have to use one more low-pass filter (for instance): the 12~kHz low-pass filter. In the following we call this mapping the audio mapping.

\pa The second new mapping we have introduced is rather uncommon but realizes an extension of the decomposition proposed by Leipp \cite{Leipp}. Leipp introduces  the 8 following subbands:  50-200 Hz (bass),  200-400 Hz (low), 400-800 Hz (low-medium), 800-1200 Hz (medium), 1200-1800 Hz (high-medium), 1800-3000 Hz (high), 3000-6000 Hz (over-high), 6000-15000 Hz (stridence). We have added two extreme bands, 0-50 Hz (low-bass) and 16000-$\frac{F_s}{2}$ Hz (high-stridence), according to our future aims. One may note that Leipp's mapping is not regular (neither linear nor logarithmic) as we have two octaves for the bass, one octave for the low, one octave for the low-medium, one fifth for the medium, one fifth for the high-medium, one sixth for the high, one octave of the over-high and on twelveth for the stridence. This mapping has been derived by Leipp by tests with musicians (whose hearing has been systematically veri\-fied using audiograms) and real stimuli: pieces of music or real voices. To determine theses subbands, called the sensible bands, Leipp plays the stimuli and from time to time introduces a rejection filter with controllable bandwidth. For each sensible band, the listening is repeated, from a non noticeable rejection to the sensible one,  by increasing the bandwitdth of the rejection until almost the subjects are able to notice the change of sound coloration. 

\pa As Leipp, we have not considered the critical bands mapping \cite{Zwicker} because such bands are derived using laboratory stimuli (sinusoids, noise) which are not real sounds. Using sinusoid is often related to the idea of the additive synthesis principle and because of the use of the impulse response or transfert functions models. We do not think the linear models are really pertinent when studying the sources or the mechanisms inside or near them, for instance. Moreover Leipp \cite{Leipp} argues that, according to the gestalt theory the "whole" is not perceptively equivalent to the sum of its parts and that to understand the influence of a component you must suppress it and then study the perception of the "whole" without this component. It seem not to be a good idea, according to the gestalt theory, to study just the isolated  component. A close idea is advanced by Blauert \cite{Blauert} when he precises that studying the perception of one source is not sufficient because some new laws are needed when there are more than one source (cf. p.  36). This principle is coherent with our tool which permits to select which components are kept and/or altered to re-synthetize the "whole". Moreover, as we also intend to study stereophonic recordings in the future, we would encounter the problem of demasking/masking with real sounds which are non coincident. This is another argument for not using the critical bands because they are determined with non-real and coincident sounds: demasking can happen when the masking and the masked sources are no more coincident. Moreover, for a first approach, managing 10 bands seems easier than 24 bands or even 25 bands. But it could be also interesting to make a comparison between sensible and critical bands. In the following, we call the sensible bands decomposition: the Leipp mapping.

\pa We must precise that the digital filters we have used are sharper (499 coefficients) than the analog ones used by Leipp (Butterworth of second order {\it a priori}). We think that Leipp used the Butterworth filter for the conservation of the energy between the bands and that a second order permits a rather easy tune of the bandwidth. We intend to study the influence of the sharpness of the separation filters in the future as our tool permits to use any kind of FIR linear phase filter with an odd number of coefficients.

\pa In the next paragraph, we propose some clues to display the information of the complete analysis.

\subsection{Ideas to display useful information}
\pa In this section we review the possible ways to display the information of the analysis first from a global point of view and then from a local point of view. By introducing these two points of view, we have chosen to follow the ideas of Leipp \cite{Leipp}:
\begin{itemize}
\item by the global point of view, we try to point out what practicians call the sound balance, or the global sonority or the global coloration of a stimulus which means that we want to get a curve giving the repartition of the total energy in the stimulus;
\item  the local point of view deals with the evolution across the time of the energy repartition like, for instance,  in a sonogram. 
\end{itemize}

\subsubsection*{The global point of view}

  \begin{figure}
\begin{center}
\includegraphics[width=7.5cm]{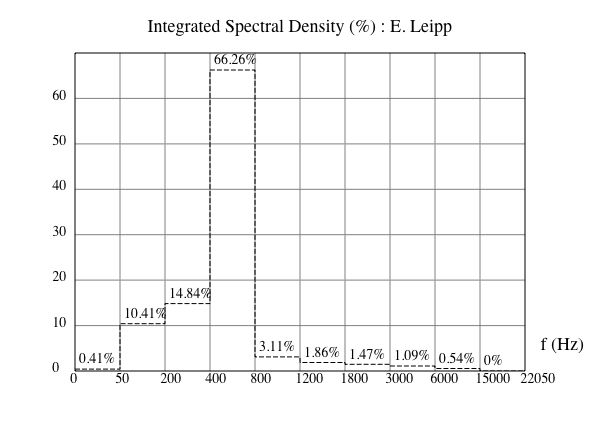} 
\includegraphics[width=7.5cm]{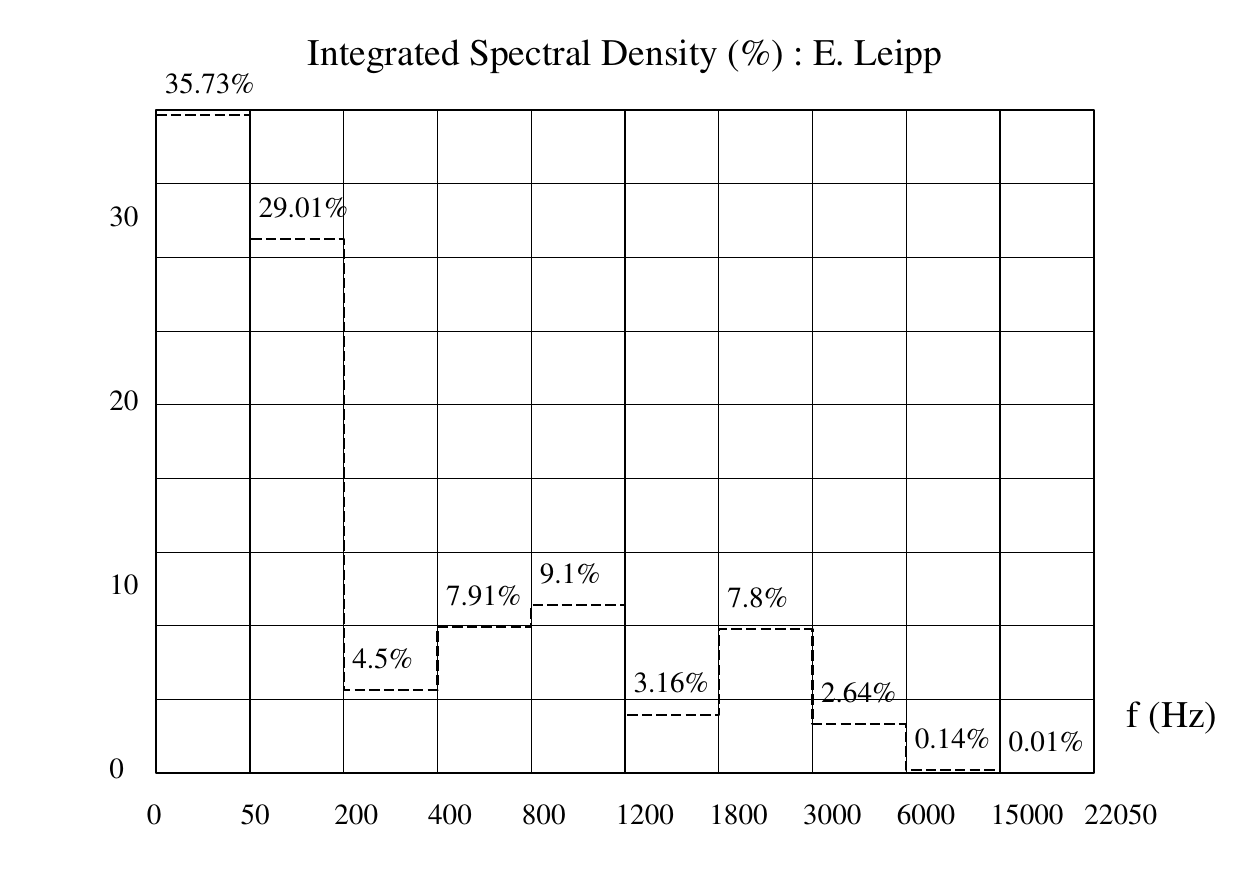} 
\caption{Plots of the ISD using the Leipp mapping with equal bandwidths plot: (left) speech, (right) pressure measurement.}
\end{center}
\end{figure} 

\pa For the global point of view and considering the little number of subbands (around 10 subbands) we have adopted the concept of the integration of spectral density (ISD) introduced by Leipp. For each subband, we calculate the local powers: the square of signal. Then we calculate the numerical integral of this power, given by the sum of all the local powers. To derive the relative balance of energy, we divide each subband integral by the sum of all the integrals and multiply each result by 100. At this time, we have for each subband a number which represent its relative weight, in percent and we can plot all the relative weights on the same figure to get the ISD. Considering the ISD, we can note that the use of sound files rather than measurements is not longer a problem because we want to study the relative balance of energy between the subbands. But using sound files permits us to listen to the subbands or to a mix of them. 

\pa With the ISD we just have the information about the relative repartition of the energy but do not have any idea of the level of the energy. To access the quantitative level of the signal energy, a measurement of the mean level (in dB SPL for instance) could be sufficient.

\pa For  a stereophonic signal, we think that the plot of one ISD by channel is the right idea. But we will have to investigate, with practicians and numerous analysis, the opportunity to calculate either two independent ISD giving the relative balance in each channel or one stereophonic ISD which means that each global subband energy is divided by the sum of all the subbands of both channels: 20 subbands in the case of the Leipp mapping for instance. This approach could be associated with the fact that we listen to only one acoustic scene even in the case of a stereophonic signal. A last question to investigate  deals with the necessity to display or not a strong visual information of the width of the subbands. In the first case we can just plot a bar plot with equal width in conjonction with the writing of the name of the bands or  by writing the limit frequencies on the x axis (see figure 8). In the second case we can plot the ISD with real bandwidths but using a logarithmic scale in basis ten (for the decades) or two (for the octaves). But in this case, it becomes difficult to distinguish the high subbands because they are too thin which explains that for the moment we have been using the first solution: plot of ISD as a bar plot with equal width.

\subsubsection*{The local point of view}

\pa To get a view of the evolution of the energy balance across the time, the idea of a spectrogram-like plot  could  be rather pertinent. Indeed, we think that the  plots of the different waveforms of the subbands (and of both plots in case of stereophonic signals) gives too many information at a same time. But it could be interesting to keep the plot of the waveforms as an option in order to be able to focus on a particular moment when necessary. This idea needs to be tested.

\pa Several representations of the energy balance seem to be interesting at this moment:
\begin{enumerate}
\item the local ISD if we plot the energy balance, expressed in percent, calculated for each sample; 
\item like in the spectrogram a plot of the temporal evolution of the ISD calculated for a sliding window (rectangular or not) of finite length to introduce a forgot factor in the analysis, with an advancement step to define;
\item the plot of the evolution of the ISD across the time which means that the normalization factor is updated at each new sample;
\item the plot of the ISD gradient given by the filtering of  the ISD by $H[z]=1-z^{-1}$.
\end{enumerate}

 \begin{figure}
\begin{center}
\includegraphics[width=7.5cm]{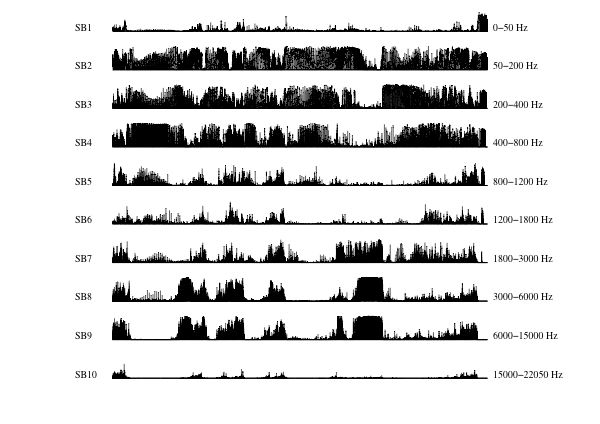} 
\includegraphics[width=7.5cm]{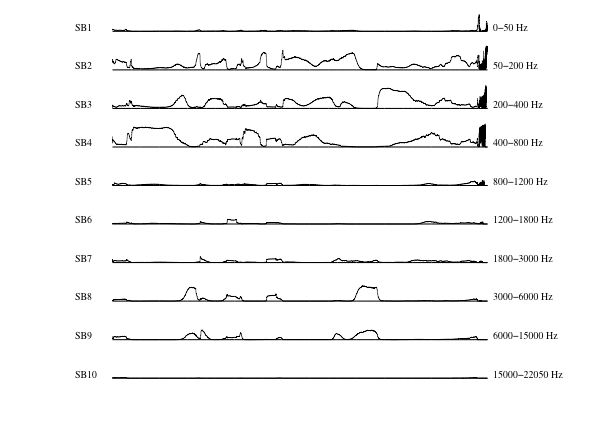} 
\includegraphics[width=7.5cm]{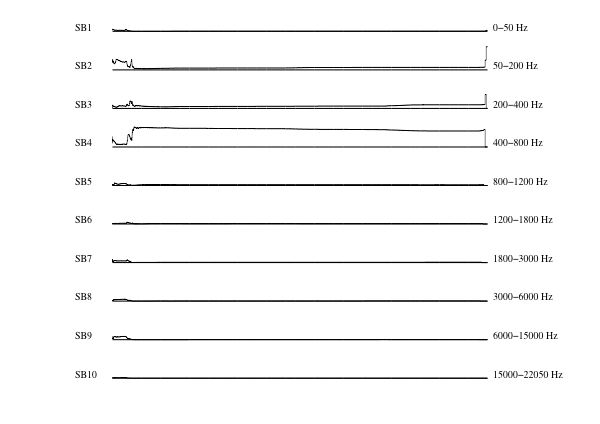} 
\includegraphics[width=7.5cm]{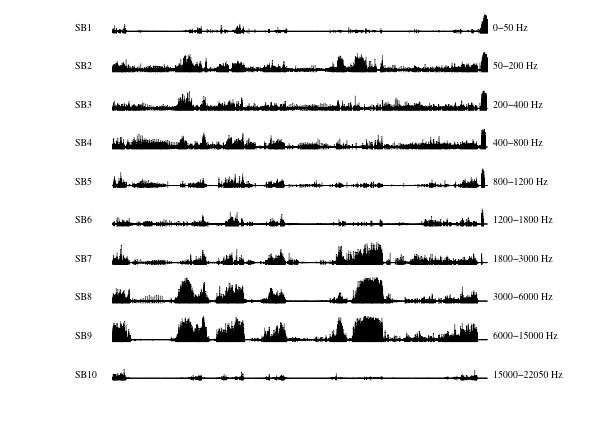} 
\caption{Plots of the TISD using the Leipp mapping for the voice signal: (top, left) sample by sample, (top, right) using a rectangular window of 2048 samples and an advancement step of 1 sample, (bottom, left) evolutive normalization factor, (bottom, right) gradient.}
\end{center}
\end{figure} 
  
  \begin{figure}
\begin{center}
\includegraphics[width=7.5cm]{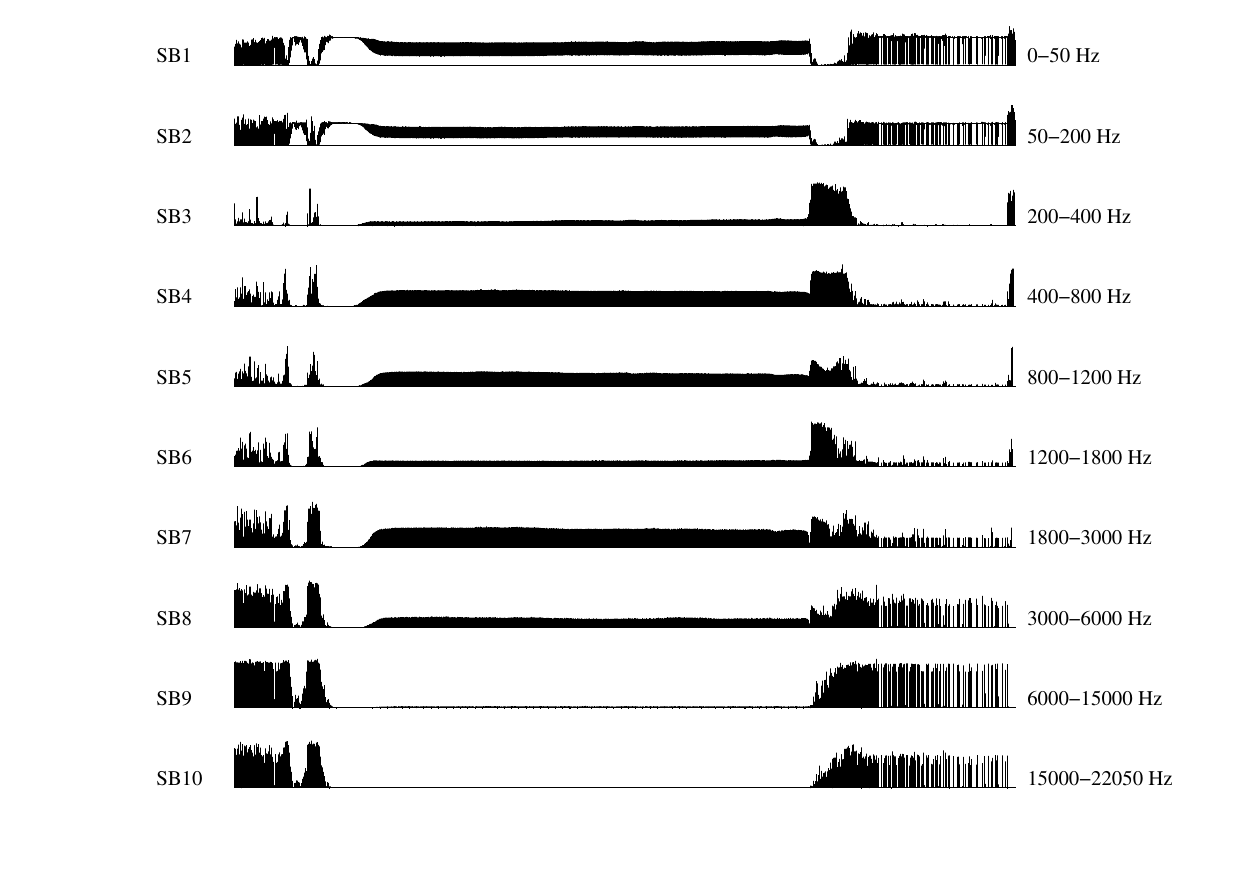} 
\includegraphics[width=7.5cm]{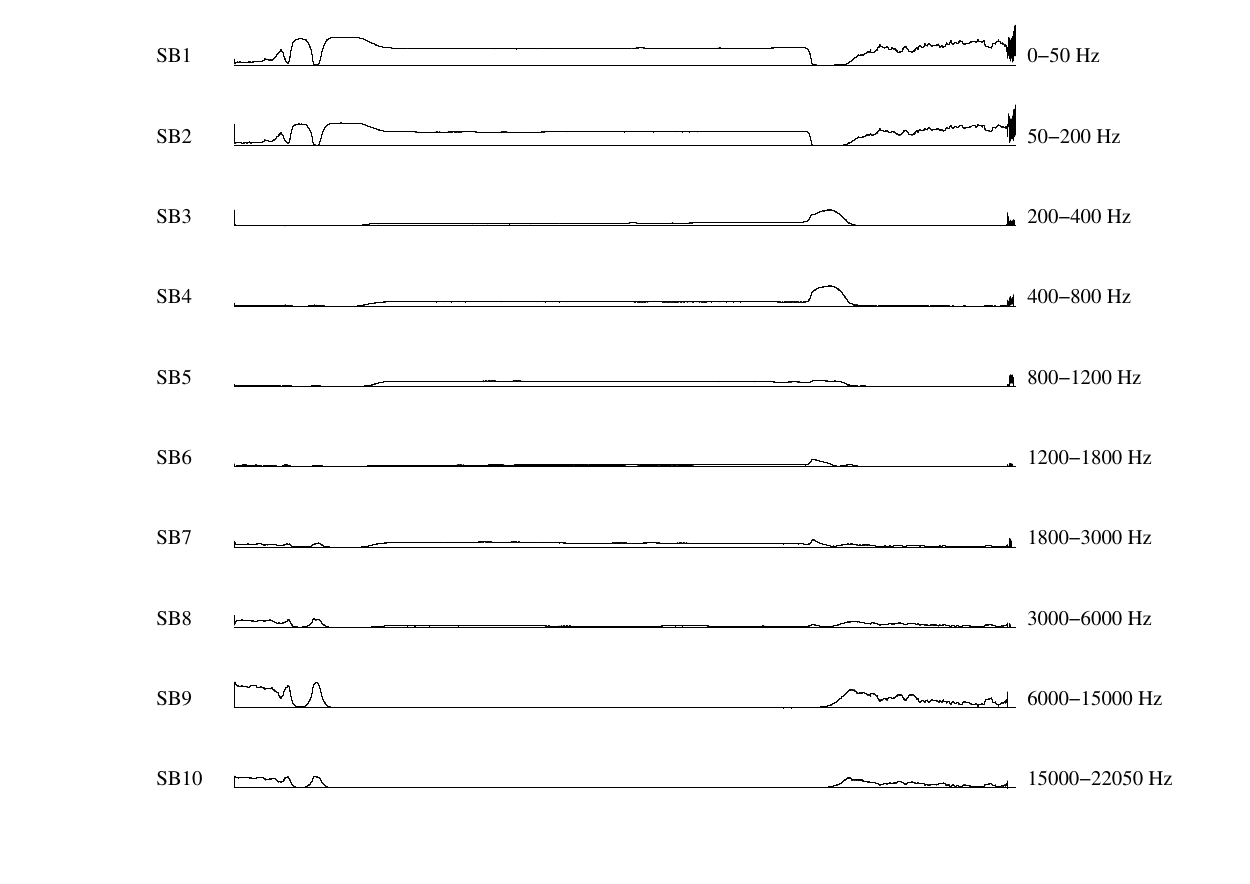} 
\includegraphics[width=7.5cm]{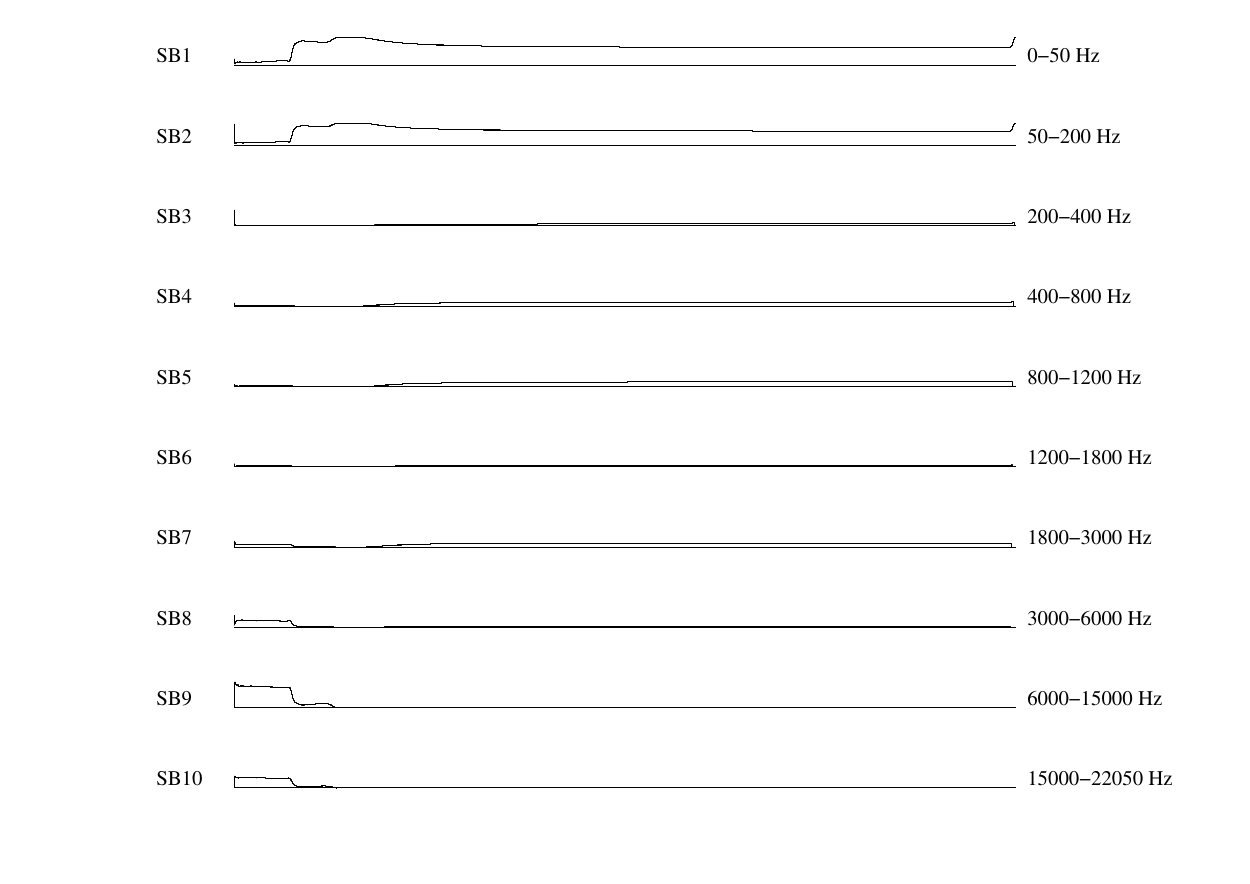} 
\includegraphics[width=7.5cm]{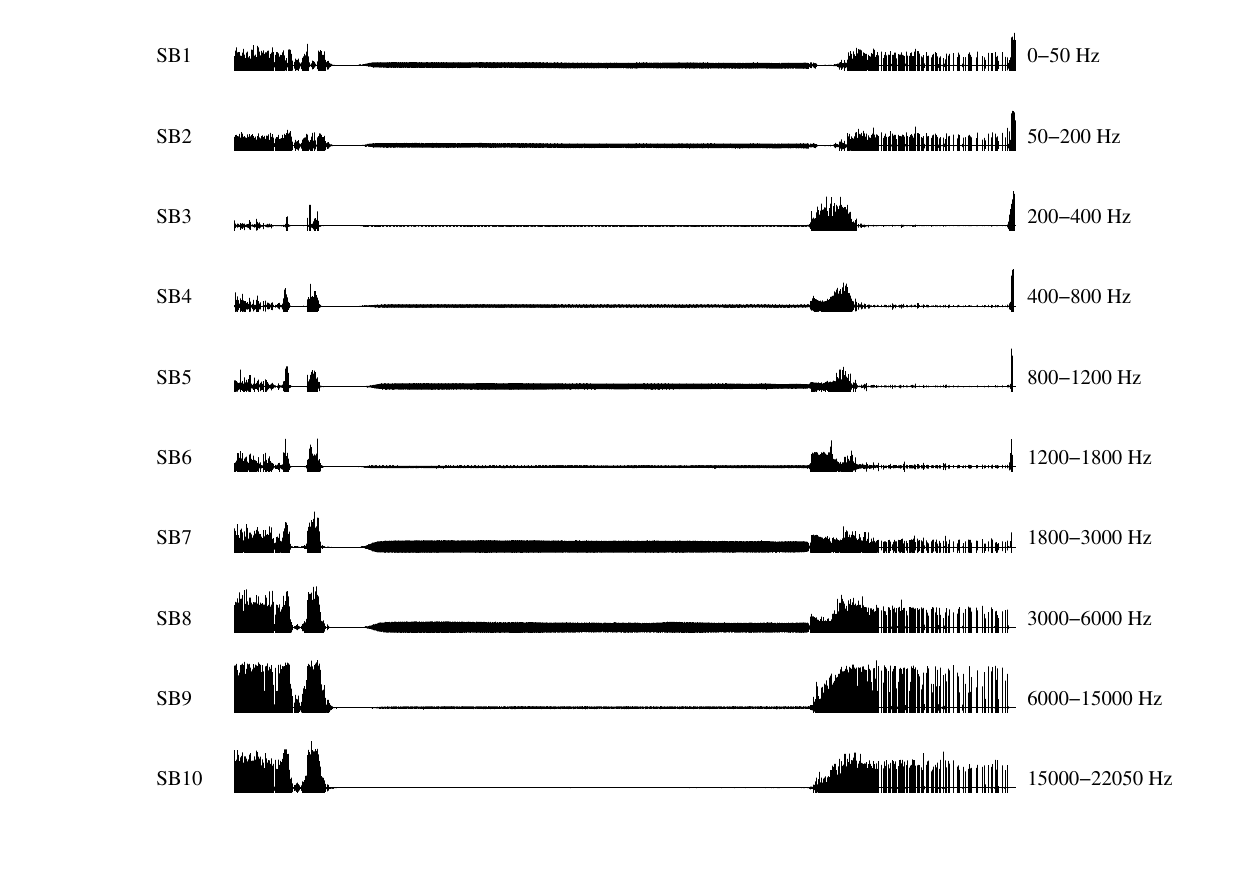} 
\caption{Plots of the TISD using the Leipp mapping for the pressure measurement: (top, left) sample by sample, (top, right) using a rectangular window of 2048 samples and an advancement step of 1 sample, (bottom, left) evolutive normalization factor, (bottom, right) gradient.}
\end{center}
\end{figure}

\pa The four representations are illustrated on figure 9 and 10 for both test signals. Leipp says that, with its analog implementation of the ISD, it is interesting to see how the ISD evolves across the time during the analysis but in his case, the analog implementation only permitted the third possibility (without drawing just seeing). We have all these four plot tools available for the moment and we think that the choice, if possible, should be done with practicians: one can imagine that different practicians  would prefer different plots; in such a case the compromise could be to let the user to configure the analysis devices. In the following we call these representations the temporal ISD (TISD).

\clearpage
\section{Discussion and perspectives}
\pa At this time, we can suggest that the first mapping to compare are the audio and Leipp ones. The audio mapping correspond to classical octaves filters easily available which can explain that sound engineers can be used to considering them are an helpful tool. It seems that such a decomposition of the frequencies axis is proposed as a first tool to my sound engineers students at Louis Lumière or at the Formation Supérieure aux Métiers du Son  (FSMS) in Paris. But this decomposition is quite regular and we are not sure it really match with the human perception. We think the so-called Leipp mapping could be quite a nice approach and should be deeply investigated. We think that some new experiments are needed to determine, with sound engineers and musicians, mainly, if we should use sharp filters and or move a little the cutting frequencies as the first ISD was built with simple analog filter and as we consider a digital version of the ISD. An alternative solution may be to propose to the end user to choose the mapping he wants. An interesting study could be to compare the relevance of the sensible (Leipp mapping) and critical bands as the first were derived using pieces of music or speech signals and the second ones using "laboratory" signals. This experiment could investigate the linear approach (assumption of elementary components) and the gestalt theory of the "whole".

\pa Some experiments are needed to determine which displaying tools could be pertinent in the context of sound production, acoustical and perception research, music practize or composition. It seems interesting to adopt the global energy balance and its evolution. We do not have a clear idea, at this point, to determine which information is relevant and useable concerning the evolution of the energy balance. To see what answer to propose we also need to work with the practicians making lots experiments. It may be also interesting to see if the real-time context will introduce the need for rather different displaying tools. It will be necessary to consider which are the displaying tools really used, for instance, by the sound engineers when they are working behind the console  and what are their needs. This interview-like process will begin soon. The question of the stereophonic case (two independent ISD for both channels or one global ISD?) and the extension to multi-channel need to be also considered.

\pa We can now introduce the intended applications for this analysis device which explain why we have begun this study on hearable sound analysis.

\pa We have been working with several students on the way to extract, characterize and/or re-synthetize  ambiance and (fixed or moving) sources from audio scenes recorded with common stereophonic technics, in the context of MPEG 4.  And these analysis tools will constitute the first tool for this study as we intend to work only with temporal indices collected in the available subbands.

\pa But, this project is also a part of a bigger project devoted to the question of Acoustics. Indeed, when considering and listening, for instance, the pressure measurement performed during a blow on a diatonic harmonica some questions appear about Acoustics. According to the reproduction device, one may be able to notice some differences while listening to the whole signal or to a mix of it without the lower subband (0-50 Hz). In some cases, the whole signal sounds louder than the mix. But, according to the classical approach of Acoustics, the 0-50 Hz (at least) correspond to the mean flow and the upper content to the acoustics: with the 0-50 Hz the signal has not a zero mean, which is one major classical assumption for Acoustics... We had worked on the way to play chromaticcally on a diatonic harmonica and the physical model we had built  \cite{Millot} does not introduce any classical equation based on propagation phenomena: there is no propagation in the approach and we did not introduce any separation between the mean flow and acoustics as we have been thinking that we have a flow which temporal variations are perceived as sound. The results for the drawn notes, bends, overblows and overdraws are all similar and the agreement between the numerical simulations and the measurements are impressive when thinking about the model. In fact, we have been dealing with the problem of sound production  inside and near the sources. Some rather similar results have been collected, for instance, for loudspeakers port design \cite{Pellerin} and should be reported in another paper at the same  AES Convention: the right way to consider the sound production and control is the study of the flow inside the port. These strange results all ask again the question of the nature of the sound: what is it, when does it begin and how can we model it with equations. Some more experiments are needed to propose a potential extension to the definition of Acoustics. For this, the analysis tools will also be used to investigate again the mechanisms of sound perception.

\pa Following Leipp \cite{Leipp}, we can already list several tests and experiments to investigate the proposed analysis tools, Acoustics and Perception:
\begin{itemize}
\item analysis of the difference of sound synthesis models of instruments and real instruments  in order to precise which are the  missing phenomena and the quality of the synthesis;
\item study of real acoustical phenomena without any "model" based {\it a priori} and with experimental devices able to catch the signals from almost 0 Hz limit;
\item comparison of several musical instruments: for instance used to play the same pieces of music by the same musicians in order to point out again \cite{Leipp} that a skilled musician can manage to play with almost the same energy balance each instrument while a beginner would not be able to do it;
\item comparison of loudspeakers using pieces of music or speech to derive subband global directivities (measurements of ISD for several position of the pressure sensor) rather than monochromatic ones, study of the low frequency components and of the properties of the sources;
\item similar comparison of different microphones to investigate their practical properties and qualities (the so-called "coloration for instance);
\item measurements of rooms responses when the excitation level varies to point out the linear or non-linear behavior of these rooms;
\item comparison of several rooms acoustics;
\item investigation of the masking processes using real sources or diffusion (monophony, stereophony, multi-channel or wave front synthesis if possible) of recorded ones: coincident sources, non-coincident sources, moving sources;
\item study of the relevance and of the eventual pedagogic interest (for sound engineer students) of the analysis tools in an esthetic context: analysis of different recordings and/or versions of given pieces of music; analysis of whole albums and/or discography or major musical bands or sound engineers or producers; analysis of representative pieces of musical of musical styles; analysis of the greatest hits over the years in several styles; analysis of student mixes;
\item investigation of the compression technics.
\end{itemize}

\section{Conclusion}
\pa We have presented and discussed several possibilities for the sound analysis: audio or Leipp mappings seem to represent a relevant first approach.  The associated synthesis  algorithm is simply a mix of the selected subbands with or without any modification. This analysis permits to listen the desired mix without any distorsion and, in fact, the algorithms are rather tunable due to the use of a modified \`{a} trous principle. Some tools for displaying the information have been introduced, which give access either to the global energy balance (the sound "coloration") or to the temporal evolution of this energy balance. Numerous applications have been suggested and these tools will be used to reinvestigate Acoustics and Perception. Experiments with practicians are needed to determine the most relevant mapping and displaying tools if possible. After (or for) this experimental phase a real-time implementation of the whole process will become a necessity.



\begin{thebibliography}{99}
\bibitem{Mallat}
S. Mallat, A wavelet tour of signal processing, second edition, Academic Press, 1999

\bibitem{Bijaoui}
A. Bijaoui, J. L. Starck and F. Murtagh, Restauration des images multi-échelles par l'algorithme à trous, Traitement du signal et des images, Vol. 11, pp. 229-243, 1994

\bibitem{Holdschneider}
M. Holdschneider, R. Kronland-Martinet, J.Morlet and Ph. Tchamitchian, {\it A real-time algorithm for signal analysis with the help of the wavelet transform}, in {\it Wavelets, time-frequency methods and phase space}, pp. 289--297, Springer-Verlag, Berlin, 1989

\bibitem{Shensa}
M. J. Shensa, {\it Discrete wavelet transforms: wedding the à trous and Mallat algorithms}, Proc. IEEE Trans. Signal Process., Vol. 40, pp. 2464--2482, 1992

\bibitem{Leipp}
E. Leipp, {\it L'intégrateur de densité spectrale {IDS} et ses applications}, in {\it Bulletin du Groupe d'Acoustique Musicale No. 94}, Laboratoire d'Acoustique Musicale, Université Paris 6, 1977

\bibitem{Zwicker}
E. Zwicker and H. Fastl, {\it Psycho-acoustics, facts and models}, second edition, Springer, 1999

\bibitem{Blauert}
J. Blauert, {\it Spatial hearing, the psychophysics of human sound localization}, revised edition, MIT Press, 2001

\bibitem{Millot}
L. Millot, {\it Free reed instruments: clues for a physical model}, in Proceedings of Stockholm Music Acoustics Conference 2003 (SMAC 03), August 6-9 2003, 329--332 

\bibitem{Pellerin}
G. Pellerin, {\it private communication}, Laboratoire d'Acoustique Musicale, Université Paris 6, 2004

\end{thebibliography}
\end{document}